\documentclass{svproc}
\usepackage{url}
\usepackage{amsmath,amsfonts,graphicx}

\newcommand{\vev}[1]{{\left< {#1} \right>}}

\newcommand{\ket}[1]{{\left| {#1} \right>}}

\begin{document}
\mainmatter              
\title{Non-invertible symmetry breaking in a frustration-free spin chain}
\titlerunning{Non-invertible symmetry breaking}  
%
\author{Akash Sinha\inst{1}, Vivek Kumar Singh\inst{2},  Pramod Padmanabhan\inst{1}, Kun Hao\inst{3,4} and Vladimir Korepin\inst{5}}
\authorrunning{Akash Sinha et al.} 
%
\tocauthor{Akash Sinha, Vivek Kumar Singh,  Pramod Padmanabhan, Kun Hao and Vladimir Korepin}
\institute{School of Basic Sciences,\\ Indian Institute of Technology, Bhubaneswar, 752050, India,\\
\email{akash25phys@gmail.com, pramod23phys@gmail.com}
\and 
Center for Quantum and Topological Systems (CQTS), NYUAD Research Institute,\\ New York University Abu Dhabi, PO Box 129188, Abu Dhabi, UAE,\\
\email{vks2024@nyu.edu}
\and
Peng Huanwu Center for Fundamental Theory, Xi'an, 710127, China,
\and
Institute of Modern Physics, \\ Northwest University, Xi'an, 710127, China,\\
\email{haoke72@163.com}
\and
C. N. Yang Institute for Theoretical Physics, \\ Stony Brook University, New York 11794, USA,\\
\email{vladimir.korepin@stonybrook.edu}}

\maketitle              

\begin{abstract}
A nearest-neighbor, frustration-free spin $\frac{1}{2}$ chain can be constructed {\it via} projectors of various ranks \'{a} la Bravyi-Gosset. We show that in the rank 1 case this system is gapped and has two ground states resembling ferromagnetic states. These states spontaneously break the non-invertible symmetry connecting them. The latter is proved using the machinery of algebraic quantum theory. The non-invertible symmetries of this system do not come from a duality. 

\keywords{Non-invertible symmetry, Spontaneous symmetry-breaking, Algebraic quantum theory}
\end{abstract}

\section{Introduction}
Wigner's theorem states that global symmetries in quantum mechanics are implemented by unitary or anti-unitary operators. In particular, they are invertible. Recently the notion of symmetry has broadened to accommodate {\it non-invertible operators} as symmetries \cite{shao 1,nameki 1}. These form algebraic structures that generalize groups. The origin of such non-invertible symmetries can be traced back to a duality of the system. The standard example being the Kramer-Wannier duality in the Ising model \cite{fendley 1,fendley 2}. We will demonstrate a different origin of non-invertible symmetries through an example of a nearest-neighbor, frustration-free spin $\frac{1}{2}$ chain with periodic boundary conditions.

 The translationally invariant Hamiltonian is
\begin{eqnarray}\label{eq:Haf}
    H_{af} = \sum\limits_{j=1}^N~h_{j,j+1}=\sum\limits_{j=1}^N~A_jF_{j+1}~;~N+1\equiv 1,
\end{eqnarray}
with $A$ and $F$ as hermitian rank $1$ projectors
\begin{equation}\label{eq:AFterm}
    A = \frac{1}{(1+a^2)}\begin{pmatrix}
        1 & a \\ a & a^2
    \end{pmatrix}~;~F=\frac{1}{(1+f^2)}\begin{pmatrix}
        1 & f \\ f & f^2
    \end{pmatrix} ; ~~ a,f\in\mathbb{R}.
\end{equation}
This system has a positive spectrum with the ground state having zero energy. There are two ground states resembling ferromagnetic states
\begin{eqnarray}
    \ket{g_a}  =  \otimes_{j=1}^N~\ket{0_a},~\ket{g_f}  =  \otimes_{j=1}^N~\ket{0_f}~;~\ket{0_x} \equiv \frac{1}{\sqrt{1+x^2}}\begin{pmatrix}
        -x \\ 1
    \end{pmatrix},
\end{eqnarray}
with $\ket{0_{a(f)}}$ being the null states of $A$($F$) respectively. These ground states are annihilated by both the global Hamiltonian $H_{af}$ and its local terms, making the system frustration-free. Note also that the ground states become orthogonal only for infinite $N$.

Theorem 1 of \cite{Bravyi2015} provides the conditions for a nearest-neighbor spin $\frac{1}{2}$ system to be gapped by studying the eigenvalues of a matrix
$$T_\psi= \begin{pmatrix} -\vev{1_a,1_f|0,1} & \vev{1_a,1_f|1,1} \\ -\vev{1_a,1_f|0,0} & \vev{1_a,1_f|1,0} 
\end{pmatrix}~;~\ket{1_x} \equiv \frac{1}{\sqrt{1+x^2}}\begin{pmatrix}
        1 \\ x
    \end{pmatrix}. $$
Here $\ket{0}$ and $\ket{1}$ form the canonical basis of $\mathbb{C}^2$.     
If the eigenvalues of $T$ have non-zero absolute value and are equal then the system is gapless. In all other situations the system is gapped. In our case these eigenvalues are 
$$\left\{0, \frac{f-a}{\left(1+a^2\right)\left(1+f^2\right)}\right\},$$
implying that $H_{af}$ is gapped. This continues to be true for the special case of $a=f$. 


\section{Spontaneous breaking of non-invertible symmetries}
The operators 
\begin{eqnarray}\label{eq:globalNIsymmetries}
    S_{f} = \prod\limits_{j=1}^N~F_j^\perp,~~S_{a} = \prod\limits_{j=1}^N~A_j^\perp,
\end{eqnarray}
are non-invertible and they commute with the Hamiltonian $H_{af}$\footnote{As they also commute with the local terms of $H_{af}$, they generate a {\it commutant algebra} which was recently introduced to study Hilbert space fragmentation \cite{sanjay 1}.}.
They act on the ground states as
\begin{eqnarray}
    S_{f,a}\ket{g_{a,f}} & = & \left(\frac{1+af}{\sqrt{1+f^2}\sqrt{1+a^2}}\right)^N~\ket{g_{f,a}},~~S_{f,a}\ket{g_{f,a}}  = \ket{g_{f,a}}.
\end{eqnarray}
In the large $N$ limit the two ground states are separated by an infinite energy barrier and so we expect them to spontaneously break the non-invertible symmetries connecting them. We will use the methods of algebraic quantum theory \cite{streater 1} to rigorously prove this.

In the thermodynamic limit the infinite lattice is rightly described by $\mathbb{Z}$, the set of all integers. A local operator is one that acts non-trivially on a finite subset of ${\mathbb Z}$. The set theoretic union of such local operators yields the local algebra ${\cal A}_{\rm loc}$. Elements of this algebra act on a finite dimensional Hilbert space as they act non-trivially only on a finite number of sites. This allows us to define an operator norm. The completion of ${\cal A}_{\rm loc}$ with regard to this norm yields the \textit{quasi-local} algebra ${\cal A}$, which we will refer to as the algebra of observables. For our system this is $\otimes~{\rm Mat}_2(\mathbb{C})$, spanned by the set 
$$ \{A_i, F_i, A_iF_i, F_iA_i\},~i\in\mathbb{Z}.$$

The non-invertible symmetries of $H_{af}$, are spontaneously broken if, in the thermodynamic limit, the two ground states belong to inequivalent representations of the algebra of observables \cite{strocchi 2}. These Hilbert spaces, ${\cal H}_a$ and ${\cal H}_f$ can be constructed by the action of the local observables on $|g_a\rangle$ and $|g_f\rangle$, respectively. They carry the representations of the algebra ${\cal A}$, which we denote as $\pi_a$ and $\pi_f$, respectively. Define the \textit{order operator}
\begin{eqnarray}
    {\cal O}=\lim_{N\uparrow \infty}\frac{1}{N}\sum_{j=1}^N\left(F+FA-A-AF\right)_j.
\end{eqnarray}
It can now be verified that
\begin{eqnarray}
    \pi_a({\cal O})=\sigma~{\rm Id}=-\pi_f({\cal O}),\quad \sigma=\frac{(a-f)^2}{{(1+a^2)(1+f^2)}}.
\end{eqnarray}
Here we made use of the fact that, since the Hilbert space ${\cal H}_a$ is constructed by the action of the local algebra on $|g_a\rangle$, it is spanned by the states which differ from $|g_a\rangle$ only on \textit{finitely} many sites. Similar arguments also hold for ${\cal H}_f$. This at once implies the absence of any intertwiner ${\cal T}$ which connects these two representations
\begin{eqnarray}
    \mathcal{T}\pi_a({\cal O})\neq\pi_f({\cal O})\mathcal{T},
\end{eqnarray}
establishing the spontaneous breakdown of the non-invertible symmetries.

\section{Outlook}
We have studied a non-invertible symmetry breaking phase in a frustration-free spin chain. It is interesting to note that the origin of this global non-invertible symmetry is different from that of the current literature on this subject. There are several directions for future work. 
\begin{enumerate}
    \item $H_{af}$ is isospectral with $H_{fa}$. This motivates the study of the phase diagram of $H=H_{af}+g~H_{fa}$ as $g$ varies. The system becomes gapless for certain values of $g$. It would be interesting to find their universality classes. The $g = -1$ Hamiltonian can also be related to alternating sign matrices with specific symmetries.
    \item At these points we can also look for Kramer-Wannier-like dualities and associated non-invertible symmetries.
    \item Non-invertible symmetries in higher rank frustration-free systems is also worth exploring as it may provide a framework for generating such models.
    \item Non-invertible symmetries are  expected to act (locally) by quantum operations \cite{tachikawa 1}. We wish to see how our model fits into this picture.
\end{enumerate}

\noindent\textbf{Acknowledgments}

We thank Paul Fendley for useful discussions.

The work of VKS is supported by ``Tamkeen under the NYU Abu Dhabi Research Institute grant CG008 and  ASPIRE Abu Dhabi under Project AARE20-336''. The work of K. H. was supported by the National Natural Science Foundation
of China (Grant Nos. 12275214, 12247103).
VK is funded by the U.S. Department of Energy, Office of Science, National Quantum Information Science Research Centers, Co-Design Center for Quantum Advantage ($C^2QA$) under Contract No. DE-SC0012704.

%
%

\end{document}